\documentclass[twocolumn,aps,prl,showpacs,floatfix]{revtex4}
\usepackage{graphicx}
\usepackage{amsmath}
\usepackage{hyperref}
\usepackage{latexsym}
\usepackage{color}
\usepackage{epsfig}


\begin{document}

\title{Interaction-driven Dynamics of $^{40}$K / $^{87}$Rb Fermi-Bose Gas Mixtures in the Large Particle Number Limit}

\author{C. Ospelkaus, S. Ospelkaus, K. Sengstock and K. Bongs}
\affiliation{Institut f\"ur Laserphysik, Luruper Chaussee 149, 22761~Hamburg / Germany}

\begin{abstract}
We have studied effects of interspecies attraction in a Fermi-Bose
mixture over a large regime of particle numbers in the $^{40}$K /
$^{87}$Rb system. We report on the observation of a mean field
driven collapse at critical particle numbers of $1.2\cdot 10^6$
$^{87}$Rb atoms in the condensate and $7.5\cdot10^5$ $^{40}$K
atoms consistent with mean field theory for a scattering length of
$a_{FB}=-281(15)\,a_0$ [S. Inouye {\it et al.}, Phys. Rev. Lett.
93, 183201 (2004)]. For large overcritical particle numbers, we
see evidence for revivals of the collapse. Part of our detailed
study of the decay dynamics and mechanisms is a measurement of the
($^{87}$Rb - $^{87}$Rb - $^{40}$K) three-body loss coefficient
$K_3=(2.8\pm1.1)\cdot10^{-28}\mathrm{cm}^6/\mathrm{s}$, which is
an important input parameter for dynamical studies of the system.

\end{abstract}


\maketitle

The recent realization of the BCS-BEC-crossover \cite{BCSBEC} in
strongly interacting dilute fermionic gases has allowed intriguing
insight into the regime connecting Bose and Fermi superfluidity.
Quantum degenerate Fermi-Bose mixtures are expected to offer an
alternative and complementary approach to Fermi superfluidity,
where the interaction between fermions is mediated by the
bosons~\cite{BIndCoop}, analogous to the role of phonons in solid
state superconductors. The realization of boson-mediated Cooper
pairing in optical lattices with tunable interactions through
hetero-nuclear Feshbach resonances would open up fascinating
perspectives. Furthermore, it has been pointed out that these
mixtures exhibit a wealth of novel quantum phases
\cite{QuantPhasMix}.

Exploring these phenomena relies on a detailed understanding of
interactions in ultracold Fermi-Bose mixtures which are
characterized by two scattering parameters: the s-wave background
scattering length for collisions between two bosons $a_{BB}$ and
the Fermi-Bose interspecies scattering length $a_{FB}$. For
repulsive interspecies interaction, the constituents may phase separate, 
while an attractive interaction creates an additional mean
field confinement and eventually renders the mixture
unstable~\cite{CollapsePhaseSepMolmer,CollapsePhaseSepRoth},
similar to the mean field collapse found in BECs with attractive
interaction~\cite{BoseCollapse}.

The prediction and first experimental confirmation of a large and
negative interspecies scattering
length~\cite{Ferrari2002a,Roati2002a} shifted the $^{40}$K /
$^{87}$Rb system into the focus of scientific interest. In
addition to the very precisely known
$a_{BB}=98.98(04)\,a_0$~\cite{RbScat}, the value of $a_{FB}$ is
of paramount importance for this system, setting maximum
achievable particle numbers but also determining correlations and
excitations. Published critical particle numbers for the occurrence of 
a mean field collapse ($N_K\approx 2\cdot 10^4$ and 
$N_{Rb}\approx 10^5$~\cite{FlorColl}) appeared to impose severe constraints 
on maximally achievable particle numbers, but at the same time seemed 
to indicate an excitingly large value of 
$a_{FB}=-395(15)\,a_0$~\cite{Modugno2003a}.
There is however an ongoing controversy in the scientific
community caused by recent reports on the observation of stable
mixtures at higher particle
numbers~\cite{Modugno2003a,Goldwin2004a} and measurements of
$a_{FB}=-250(30)\,a_0$~\cite{Goldwin2004a} using thermal
relaxation and $a_{FB}=-281(15)\,a_0$~\cite{JILAFeshKRb} using
Feshbach spectroscopy methods. A detailed exploration of the
regime of mean field instability is not only interesting in itself
due to the its strong dependence on the interaction parameters and
due to the dynamical behavior when the instability occurs. More
generally, it is a fundamental prerequisite for a full
understanding of $^{40}$K / $^{87}$Rb-mixtures and the controlled
achievement of fascinating physics in the regime of large particle
numbers and strong interactions, e.g. the realization of mixed
bright solitons~\cite{KarpiSol} and the observation of Fermi-Bose
correlations up to novel BCS phases in optical lattices.

In this letter, we report on reaching the to our knowledge so far
highest particle numbers in the $^{40}$K / $^{87}$Rb system only
limited by the onset of instability, which we find consistent with
$a_{FB}\approx -281\,a_0$~\cite{JILAFeshKRb}. Right at the phase
transition point of a large bosonic thermal cloud ($n\approx
4\cdot10^{14}\,\mathrm{cm}^{-3}$), we observe a strong dynamical
modification of the in-trap fermionic distribution due to the
mean-field interaction with the bosons. At even higher densities,
we observe enhanced localized loss processes in the overlap region
of the BEC and the Fermi gas, which we ascribe to a mean field
collapse based on lifetime and dynamical arguments as well as
comparison to theory. After the collapse, we observe a peaked
density distribution of the Fermi cloud subjected to the strong
localized mean field potential of the condensate. In addition, we
present a novel measurement of the three-body inelastic K-Rb decay
constant differing by an order of magnitude from the value given
in~\cite{FlorColl}.

Our setup is based on a two-species 2D/3D-MOT
geometry~\cite{Dieck2DMOT}, where the $^{40}$K vapor is produced
using self-made enriched $^{40}$K dispensers~\cite{EnrDispJILA}.
In the 3D-MOT, we achieve $1\cdot10^{10}$ $^{87}$Rb and
$2\cdot10^8$ fermionic $^{40}$K atoms within 10 seconds. The
pre-cooled mixture is loaded into a Ioffe-Pritchard type magnetic
trap (cloverleaf / 4D-hybrid, $\nu_{\mathrm{ax}}^{\mathrm{Rb}}=
11.3\,\mathrm{Hz}$ and $\nu_{\mathrm{rad}}^{\mathrm{Rb}}=
257\,\mathrm{Hz}$, i.e. $\nu_{\mathrm{ax}}^{\mathrm{K}}=
16.6\,\mathrm{Hz}$ and $\nu_{\mathrm{ax}}^{\mathrm{K}}=
378\,\mathrm{Hz}$) for RF-induced sympathetic cooling of $^{40}$K
in the $|F=9/2,m_F=9/2>$ state with $^{87}$Rb in the $|F=2,m_F=2>$
state. If all of the $^{87}$Rb atoms are removed, we currently
obtain Fermi gases of up to $3\cdot 10^6$ $^{40}$K atoms at
$T/T_F=0.2$ or $9\cdot 10^5$ at $T/T_F=0.1$.

When the $^{87}$Rb atoms are not completely evaporated, various
regimes of mixtures are accessible, ranging from dense thermal
$^{87}$Rb clouds of $10^7$ $^{87}$Rb atoms right at the phase
transition point interacting with a moderately degenerate Fermi
gas of $2\cdot 10^6$ $^{40}$K atoms to deeply degenerate mixtures
with almost pure condensates. We achieve $>1\cdot 10^6$ atoms in
the condensate coexisting with $7.5\cdot 10^5$ $^{40}$K atoms,
limited by the onset of the collapse.

\begin{figure}[tbp]
\begin{centering}
\leavevmode
\resizebox*{1\columnwidth}{!}{\includegraphics{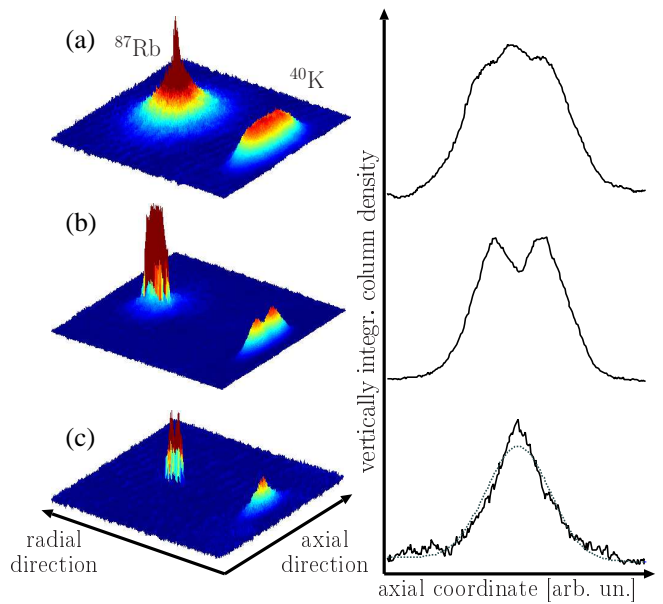}}
\end{centering}
\caption{Typical evolution of an overcritical mixture after an
evaporation ramp (rate -1\,MHz/s) stopped and held fixed for
15\,ms at 80\,kHz (a), 50\,kHz (b) and 20\,kHz (c) above the
$^{87}$Rb trap bottom of 490\,kHz. Left-hand side: 3D
representation of absorption images with false-color coding of the
optical density. $^{87}$Rb and $^{40}$K images are taken in the
same run, although at different TOF: $20 \mathrm{ms}$ ($^{87}$Rb)
and $3-5\mathrm{ms}$ ($^{40}$K). Right-hand side: corresponding
axial line profiles integrated along the vertical direction.}
\label{fig_profile}
\end{figure}

\begin{figure}[tbp]
{\centering \resizebox*{1\columnwidth}{!}{\includegraphics{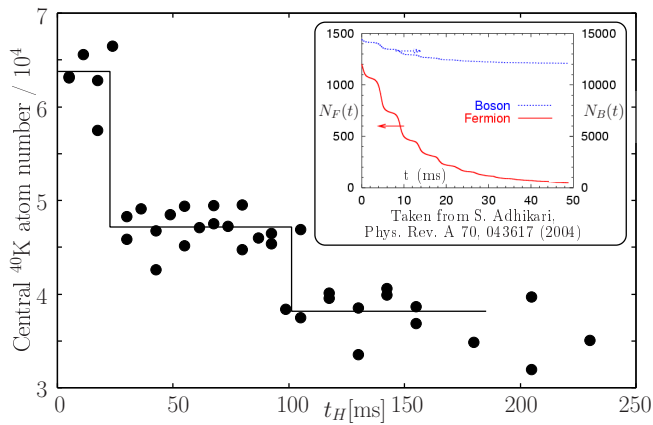}} }
\caption{Decay of the fermionic component in a slightly overcritical mixture. The first sudden drop
is the initial collapse; the second drop is one ``revival'' of the collapse.
The line is to guide the eye. For comparison, the inset shows results from
dynamical modelling of the collapse.}
\label{fig_steps}
\end{figure}

In a first set of experiments, shown in Fig.~\ref{fig_profile}, we
analyze the behavior of the mixture in these regimes during the
evaporation ramp, similar to the procedure in~\cite{FlorColl}. In
our case the dynamics becomes clearly visible in the axial
$^{40}$K density distribution, which due to the relatively short
time of flight (TOF) should closely reflect the in trap
distribution. The fast ramp speed ($1\,\mathrm{MHz}/\mathrm{s}$)
in this experiment means that we are potentially dealing with
strongly out of equilibrium samples. Already at the phase
transition point (a) of the bosonic cloud, we observe strong
distortions of the axial $^{40}$K density profile. The profile
looks like a chopped off Fermi profile with a peak in the center
of the flat top which we ascribe to the interaction with the
bosonic component (peak density $\approx
4\cdot10^{14}\,\mathrm{cm}^{-3}$). For a large BEC (b) the
$^{40}$K profile exhibits a pronounced hole in the trap center
which we ascribe to a strong localized loss process due to the
interaction with the BEC. These losses are too fast for transport
in the slow axial direction to be able to continuously maintain
the undisturbed density profile by refilling the center of the
trap from the outer regions. As we shall see, this fast loss (e.
g. from Fig.~\ref{fig_steps} $\le10\,\mathrm{ms}$) is a signature
of the mean field collapse of the mixture. After the collapse, we
observe that $^{40}$K distributions sharply peaked in the center
remain stable for relatively long timescales exceeding
$100\,\mathrm{ms}$ (c). A Fermi-Dirac fit to the data is shown as
a dotted line for comparison. We attribute this peaked
distribution to the Fermi-Bose attraction creating an additional
trapping potential for the fermions (``mean field dimple'' in the
magnetic trap potential).

The strong depletion of the Fermi cloud in the center is
accompanied by rapid particle loss of approximately two thirds of
the fermionic particle number. We ascribe this loss to the mean
field collapse of the mixture. Beyond critical particle numbers,
the mean field potential is no longer balanced by the repulsive
interaction in the $^{87}$Rb BEC and the outward bound Fermi
pressure, so that the part of the mixture overlapping with the BEC
contracts rapidly to such large densities that enormous 3-body
losses reduce the overall particle number in this region to an
undercritical value. It is therefore natural that we observe the
collapse in the vicinity of strong contracting mean field effects
of the Bose gas on the fermionic distribution (see
Fig.~\ref{fig_profile}a, c; cf.~\cite{FlorBimod}). Although the
ensemble is out of equilibrium and shows very complex dynamics
{\it after} this rapid loss~\cite{AdhCollDyn}, one can intuitively
imagine that the Fermi distribution depleted in the center of the
cloud is refilled from the outside parts of the sample on a
timescale related to the axial trap frequency, possibly leading to
repeated local collapses, until the mixture will become
undercritical and reestablish an equilibrium situation. In order
to observe this phenomenon we have prepared a mixture where the
bosonic part ($\approx 10^7$ atoms) has only started condensing
and observed the evolution of the ensemble at constant evaporation
frequency. As the condensate grows due to the mild remaining
evaporative cooling, it reaches the critical particle number of
$N_B=1.2\cdot10^6$. Due to the near-equilibrium situation before,
the collapse now only leads to a relatively small loss in total
particle number but is still clearly visible in the $^{40}$K atom
number integrated over the central part of the TOF image as shown
in Fig.~\ref{fig_steps}. After the first collapse this number
remains constant for some time and then drops abruptly
again~\footnote{The refilling is not visible for the chosen
integration area.}. Such ``revivals'' of the collapse have been
predicted in a recent numerical analysis of the collapse dynamics
(\cite{AdhCollDyn}, see inset in Fig.~\ref{fig_steps}), although
in a spherically symmetric configuration with
$\nu_{\mathrm{Rb}}=100\,\mathrm{Hz}$.

\begin{figure}[tbp]
{\centering
\resizebox*{1\columnwidth}{!}{\includegraphics{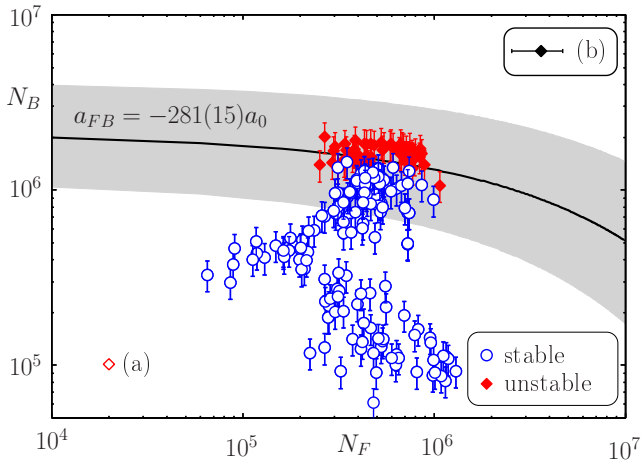}} }
\caption{Stability diagram for the $^{40}$K / $^{87}$Rb mixture.
The uncertainty in $N_{\mathrm{Rb}}$/$N_{\mathrm{K}}$ is assumed
to be 20\%/30\%. The error on $N_F$ is not critical and is given
exemplarily in (b). The solid black line is based on the theory of
ref. \cite{ChuiColl} and $a_{FB}=-281(15)a_0$~\cite{JILAFeshKRb}.
(a) is the critical particle number reported in
ref.~\cite{FlorColl} (here the trap had a different aspect ratio,
but a similar mean trapping frequency $\bar \nu_K =134\,$Hz, as
compared to our experiment with $\bar \nu_K = 133\,$Hz)}
\label{fig_stability}
\end{figure}

Analysis of similar decay series in various particle number
regimes enables us to extract a value for the critical particle
numbers for the onset of the collapse. Our findings are summarized
in Fig. \ref{fig_stability}. Situations where the decay is
compatible with 3-body decay of an undisturbed density
distribution are identified as stable. If the loss is incompatible
with 3-body decay, the situations are identified as unstable. Note
that the time scales for the two situations are clearly different,
varying from a few 100\,ms for the stable cases to about 20\,ms
for the unstable cases. The solid line in Fig. \ref{fig_stability}
is the theoretical stability limit for our trapping conditions,
based on~\cite{ChuiColl} and the following approximation to the
fermionic chemical potential:
$\mu_F=\mu_{F,0}-(g_{FB}/g_{BB})\cdot\mu_{B,0}$ where $\mu_{F,0}$
and $\mu_{B,0}$ are the Fermion / Boson non-interacting chemical
potentials and $g_{FB}$ and $g_{BB}$ are the standard coupling
constants \cite{JonPhD,QuantPhasOptLatt}. The curve is plotted for
the value $a_{FB}=-281\,a_0$; a confidence interval for the
stability border based on the uncertainty of $15 a_0$ on
$a_{FB}$~\cite{JILAFeshKRb} is indicated in grey. Our findings for
the critical particle numbers are clearly consistent with
mean-field theory of the collapse based on this value for $a_{FB}$
\footnote{Using~\cite{ChuiColl}, our critical particle numbers
yield $a_{FB}=-284\,a_0$; using~\cite{KarpiColl}, we obtain
$a_{FB}=-294\,a_0$.}.

\begin{figure}[tbp]
\begin{centering}
\leavevmode
\resizebox*{1\columnwidth}{!}{\includegraphics{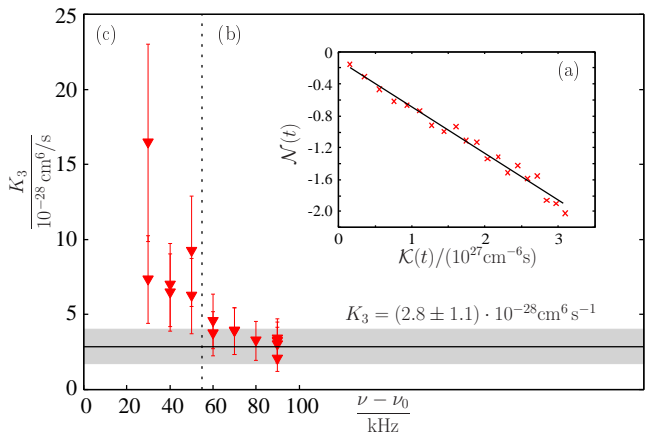}}
\end{centering}
\caption{Extraction of 3-body inelastic loss rate. (a) decay
analysis of a thermal sample, where
$\mathcal{N}(t)=\ln{\frac{N_F(t)}{N_F(0)}}+\frac{t}{\tau }$ and
$\mathcal{K}(t)=N_F^{-1}\cdot\int_0^{t}{dt' \int{d^3r
n_b^2(r,t')\cdot n_F(r,t')}}$ (b) $K_3$ values extracted in the
moderately degenerate and stable regime with peak BEC densities of
up to $4\cdot10^{14}\,\mathrm{cm}^{-3}$ are compatible with the
thermal $K_3$ average (solid line). (c) For low evaporation end
frequencies and BEC densities of
$\approx5-7\cdot10^{14}\,\mathrm{cm}^{-3}$, samples are unstable
with respect to the collapse. } \label{fig_3body}
\end{figure}

Three-body decay is the underlying loss mechanism for the collapse
of the mixture. The value of the associated rate coefficient is
therefore an important input parameter for detailed understanding
of the dynamical behavior of the collapse
(``revivals'')~\cite{AdhCollDyn}. In order to minimize
experimental uncertainties, we extract the value for the 3-body
rate coefficient from the decay of {\it non-degenerate thermal
ensembles} in an equilibrium situation over a large range of
densities. Since collisions involving two identical fermions are
strongly suppressed due to the Pauli exclusion
principle~\cite{Petrov2003a,Incao2005a}, we consider only
inelastic collisions between two bosons and one fermion. For pure
3-body loss, the decay process is then characterized by the
following rate equation:
$$\frac{\dot{N}_F(t)}{N_F(t)}=-\tau^{-1}-K_3\cdot\int\,\mathrm{d^3}r\,n_B^2(r,t)\,\frac{n_F(r,t)}{N_F(t)}$$
where $n_B$ and $n_F$ are the respective densities, $K_3$ is the
rate coefficient for three-body loss and $\tau$ the background
collisional lifetime. Integrating this equation over time yields
$$\ln \frac{N_F(t)}{N_F(0)}+\frac t\tau=-K_3\cdot\int_0^{t}\mathrm{d}t'\int\mathrm{d^3}r\,n_B^2(r,t')\cdot \frac{n_F(r,t')}{N_F(t')}$$
Based on the measured background scattering rate, the left-hand
side expression can be evaluated at any given time $T$ for a decay
series. The right-hand side inner integral is evaluated using an
equilibrium Fermi-Dirac profile for $^{40}$K and a Bose-enhanced
thermal profile for $^{87}$Rb at different time
values $t_i$. The density distributions are based on the measured
temperatures and total atom numbers at $t_i$. The outer integral
is evaluated by using a piecewise linear interpolation between the
discrete values of the inner integral. Plotting the right-hand
side as a function of the double integral, we obtain $K_3$ as the
slope of the linear plot (see inset (a) in figure
\ref{fig_3body}). Note that our measurement does not depend on the
potassium atom number calibration. As far as the $^{87}$Rb atom
number calibration is concerned, we reproduce the measurement of
$^{87}$Rb 3-body decay from ref. \cite{Rb3bodyDal}. We find a
value of $K_3=(2.8\pm1.1)\cdot10^{-28}\,\mathrm{cm}^6/\mathrm{s}$
\footnote{A 2-body decay analysis leads to a large
density-dependent variation (by a factor of 40) of the
$K_2$-values, indicating that the decay is predominantly 3-body.},
nearly an order of magnitude lower than the value found
in~\cite{FlorColl}.

The additional analysis of decay in degenerate samples at higher
densities reproduces this result as long as the mixture is stable
(\ref{fig_3body}b), but shows significant deviations for unstable
mixtures (\ref{fig_3body}c). In the regime of instability, our
analysis of the loss process is no longer valid. Due to the
strongly depleted $^{40}$K density distribution, our analysis,
relying on undisturbed density distributions, overestimates the
density overlap integrals which should lead to a too low value of
$K_3$ as extracted from the decay. Instead, as can be seen from
Fig. \ref{fig_3body}c, $K_3$ increases below evaporation end
frequencies of 60 kHz (high initial densities), proving that the
underlying very fast loss mechanism is due to the collapse.

In conclusion, we have presented measurements covering the so far
highest particle numbers available in quantum degenerate
$^{40}$K-$^{87}$Rb Fermi-Bose mixtures leading from stability to
collapse. Our measurements of critical particle numbers are
consistent with a mean field model of the collapse based on the
value of the scattering length $a_{FB}=-281(15)\,a_0$ obtained
from Feshbach resonance data~\cite{JILAFeshKRb}. This settles the
clear discrepancy between the theoretical description of the
collapse based on the above interspecies scattering parameter and
the critical particle numbers for the  mean field collapse
reported in an earlier experiment~\cite{FlorColl} (see Fig.
\ref{fig_stability}a) and indicates that these findings might be
influenced by some additional interesting effect. Part of our
detailed study of the mixture is a quantitative analysis of
three-body loss leading to an order of magnitude lower value than
in~\cite{FlorColl}. Our findings indicate that a wide range of
filling factors in large-volume optical lattices ranging in principle 
up to 2 are possible for the $^{40}$K-$^{87}$Rb system. Progress 
towards loading the mixture into a 3D optical lattice is
currently under way in several labs. Finally, reaching the regime
of mean field collapse also lays the ground for the realization of
bright solitonlike structures in a quasi 1D geometry
\cite{KarpiSol}.

\begin{acknowledgments}
We thank K. Hofmann and B. Albert for help with enriched dispenser
production. We acknowledge discussions with M. Brewczyk, M. Gajda,
K. Rz{\c a}{\.z}ewski and J. M. Goldwin as well as contributions
by R. Dinter, P. Ernst, J. Fuchs, M. Nakat, M. Succo, O. Wille and
funding by DFG under SPP 1116.
\end{acknowledgments}

\end{document}